\documentclass[a4paper,12pt]{article}
\usepackage[top=1.25in, bottom=1.25in, left=1.05in, right=1.05in]{geometry}
\usepackage[utf8]{inputenc}
\usepackage[T1]{fontenc}
\usepackage[english]{babel}

\usepackage{natbib}
\usepackage[section]{placeins}
\usepackage{setspace}

\usepackage{enumitem}
\setlist{noitemsep}  

\usepackage{amscd,amsmath,amssymb,amsfonts,amsthm,bbm,bm,latexsym,mathrsfs,mathtools,bigints}
\usepackage[subnum]{cases}	
\usepackage{dsfont}
\makeatletter
\newcommand*{\distas}[1]{\mathbin{\overset{#1}{\kern\z@\sim}}}	
\makeatother

\allowdisplaybreaks

\newtheorem{definition}{Definition}[section]
\theoremstyle{remark}

\theoremstyle{plain}

\RequirePackage[colorlinks=true, citecolor=blue, urlcolor=blue]{hyperref}

\usepackage[dvipsnames]{xcolor}
\usepackage{subfigure,epsfig,epstopdf,rotating,graphics,graphicx} 
\graphicspath{{./figures/}}
\usepackage[width=0.9\textwidth, font={footnotesize}]{caption}

\usepackage{rotating,lscape,pdflscape}
\usepackage{booktabs,longtable,float,array,multirow,colortbl,adjustbox}

\usepackage[colorinlistoftodos,color=orange!60]{todonotes}
\usepackage{regexpatch}
\makeatletter
\xpatchcmd{\@todo}{\setkeys{todonotes}{#1}}{\setkeys{todonotes}{inline,#1}}{}{}
\makeatother

\makeatletter

\def \bq{\mathbf{q}}

\def \by{\mathbf{y}}
\def \bz{\mathbf{z}}
\def \bB{\mathbf{B}}
\def \bQ{\mathbf{Q}}

\def \bZ{\mathbf{Z}}
\def \bbeta    {\boldsymbol{\beta}}

\def \bepsilon {\boldsymbol{\epsilon}}

\def \I {\mathbb{I}}
\def \R {\mathds{R}}

\def \QS {\text{$\mathcal{QS}$}}

\makeatother

\title{\vspace{-60pt} \textbf{Money Growth and Inflation: A Quantile Sensitivity Approach}
\thanks{The authors gratefully acknowledge Efrem Castelnuovo, Todd Clark, Catalina Martínez Hernández and the participants at seminars at the University of Milan and at the 6th Annual Workshop on Financial Econometrics at \"Orebro. Luca Rossini acknowledges financial support from the Italian Ministry MIUR under the PRIN project Modelling Non-standard data and Extremes in Multivariate Environmental Time series (MNEMET) (grant 20223CEZSR). The paper previously circulated under the name ``The Distributional Impact of Money Growth and Inflation Disaggregates: A Quantile Sensitivity Analysis''}
}

\author{
Matteo Iacopini\thanks{Queen Mary University of London, United Kingdom. \color{blue}\texttt{m.iacopini@qmul.ac.uk}}
\and
Aubrey Poon\thanks{\"Orebro University, Sweden and University of Kent, United Kingdom. \color{blue}\texttt{a.poon@kent.ac.uk}}
\and
Luca Rossini\thanks{University of Milan, Italy and Fondazione Eni Enrico Mattei. \color{blue}\texttt{luca.rossini@unimi.it}}
\and
Dan Zhu\thanks{Monash University, Australia. \color{blue}\texttt{dan.zhu@monash.edu}}
}

\date{\today}

\begin{document}

\maketitle

\begin{abstract}
An innovative method is proposed to construct a quantile dependence system for inflation and money growth.
By considering all quantiles and leveraging a novel notion of quantile sensitivity, the method allows the assessment of changes in the entire distribution of a variable of interest in response to a perturbation in another variable's quantile.
The construction of this relationship is demonstrated through a system of linear quantile regressions. Then, the proposed framework is exploited to examine the distributional effects of money growth on the distributions of inflation and its disaggregate measures in the United States and the Euro area.
The empirical analysis uncovers significant impacts of the upper quantile of the money growth distribution on the distribution of inflation and its disaggregate measures. Conversely, the lower and median quantiles of the money growth distribution are found to have a negligible influence.
Finally, this distributional impact exhibits variation over time in both the United States and the Euro area.


\end{abstract}


\clearpage

\doublespacing

\section{Introduction}

As stated by Isabel Schnabel, Member of the Executive Board of the ECB, in her Th\"{u}nen Lecture at the annual conference of the Verein f\"{u}r Socialpolitik: \textit{``In the two and a half years following the outbreak of the pandemic, the sum of currency in circulation and overnight bank deposits in the euro area, referred to as M1, increased by over 30\%. Over the same period, inflation accelerated from 1.2\% to 9.1\%. It peaked at 10.6\% in October 2022''}. In detail, she discusses the role of money in explaining the inflation surge and concludes that money growth is still important in unstable conditions when adverse cost-push shocks risk lifting inflation away from the central bank’s target.
Besides, the relationship between inflation and money growth has garnered considerable attention for a long time in the field of empirical macroeconomics \citep[e.g., see][]{lucas1980two,grauwe2005inflation, gertler2018monetary, benati2009long}. This relationship was particularly crucial during the inflationary period of the 1970s and 1980s. However, adopting inflation targeting in most advanced economies during the 1990s weakened this relationship due to sustained periods of low and stable inflation. The recent resurgence of inflation and the expansion of the money stock during the COVID-19 pandemic has revitalized interest in investigating this relationship among policymakers and researchers \citep[see][]{laidler2021personal,king2022monetary,borio2023does}.
Consequently, this study examines the short-run distributional impact of money growth on inflation and its disaggregate measures in the US and the Euro Area. We differ from \cite{sargent2008monetary} that refer to the theory of money in the long run and filter the data to extract the long-run component of growth rates.


This article also investigates whether there is evidence of temporal variation in the short-run distributional impact of money growth on inflation. To the best of our knowledge, our study is the first to explore the relationship between inflation and money growth in a distributional context. Previous studies have solely examined this relationship through the perspective of the conditional mean of the distribution and have led to conflicting findings \citep[see][]{grauwe2005inflation,gertler2018monetary}. The main advantage of our study is that we can directly investigate the effects of excessive money growth (or the upper quantile of the money growth distribution) on the distribution of inflation.

Indeed, the standard approach for estimating dependence between two random variables involves fitting a linear model to the data and examining the relationship between the predictor and response variables. Despite its popularity, this framework inherently models only the conditional mean of the response variable. Still, it fails to uncover the intricate dependencies within various regions of the joint distribution. Consequently, it is a somewhat limited approach to describe the dependence between the two variables.
When dependence beyond the conditional mean is of interest, a more general approach relies on quantile regression \citep{koenker1978regression}, which investigates the effects of a variable on the entirety of the distribution of the response.


In recent years, there have been significant advancements in economics exploring the application of alternative dependence measures. A notable proposal in this regard is the Conditional Value-at-Risk \citep[CoVaR, see][]{tobias2016covar}, which utilizes a system of quantile regressions to assess the escalation of tail co-movement among financial institutions. For example, these measures capture the change in the financial system's 1\% quantile when a specific institution experiences its 1\% quantile. However, the linear construction behind the CoVaR results in the unfortunate limitation that the percentage point change in response to the 1\% and 99\% quantiles is the same.

To overcome this limitation, we propose an alternative method to constructing a quantile dependence system for inflation and money growth. Our approach assumes that with a sufficiently long time series of lagged past observations, denoted as $\bz_t$, we can reliably estimate the system of conditional quantiles for both inflation and money growth, which become functional representations of $\bz_t$.
Therefore, by employing linear projections, we can determine the sensitivity of inflation quantiles to changes in the quantiles of money growth. This allows us to capture the intricate relationship between inflation and money growth across different quantiles.

In contrast to the CoVaR approach, it is important to note that the responsiveness of inflation to the 10\% money growth quantile might differ from its responsiveness to the 90\% money growth quantile. The proposed approach recognizes and accounts for the potential variation in the impact of different quantiles of money growth on inflation. This allows for a more nuanced understanding of how inflation responds to varying levels of money growth throughout the distribution.

Our empirical analysis reveals a significant positive influence of the upper quantiles of the money growth distribution on the distribution of inflation and its disaggregate measures in the US and the Euro area. Specifically, we find evidence of an increasing positive skewness in the overall distribution of US Personal Consumption Expenditure (PCE) inflation and Euro area Harmonised Index of Consumer Prices (HICP) inflation. This upward skewness primarily stems from a shift in the distribution of services inflation towards higher values.
In contrast, the distributional changes caused by perturbations of the lower quantiles of money growth are not significant.

Moreover, our analysis uncovers that, at the low to median quantiles of the money growth distribution, money growth has a relatively negligible distributional effect on inflation. Only in the upper tails of the money growth distribution do we observe a substantial impact on the overall distribution of inflation. Thus, our results suggest that the relationship between inflation and money growth is weak during periods of economic stability and low volatility, as indicated by the median of the money growth distribution. It is only during exceptional events, such as the high inflation period after the pandemic (represented as the upper quantiles of the money growth distribution), that the relationship between inflation and money growth becomes statistically significant.

This finding aligns with \cite{golosov2007menu}, underscoring the significant role of menu costs in maintaining price stability over time. It highlights that price adjustments by firms typically occur in response to notable idiosyncratic shocks in the economy, allowing prices to remain relatively constant otherwise. 
Moreover, our findings agree with \cite{Schnabel2023}, indicating that the influence of monetary expansion on inflation is contingent upon economic circumstances. Besides, we support her assertions, revealing that within the upper quantile of money growth, the distribution of inflation skews towards a more positive trajectory.
Our discovery is consistent with \cite{blanco2022nonlinear}, demonstrating that substantial monetary shocks result in more than a doubling effect on the frequency of price changes within the economy. Overall, our empirical findings emphasize the necessity of incorporating the distributional effects of money growth into future structural macroeconomic models.

This article is organized as follows.
Section~\ref{sec:methods} introduces and describes the quantile dependence framework. Section~\ref{sec:results} details the empirical application undertaken in the paper and presents the results derived from our empirical analysis. Finally, Section~\ref{sec:conclusion} concludes.

\section{Methodology}  \label{sec:methods}

This section presents a framework for systematically analyzing the impact of a perturbation in the conditional quantile of one variable on the conditional quantile of another variable in the system. This analysis can be effectively and meaningfully conducted when both of these variables are conditioned on the same set of covariates. 
The proposed method initially involves constructing a system of conditional quantiles for both variables using the standard quantile regression method. Subsequently, a direct relationship between the quantiles of the two variables is established through linear projection, and the perturbation analysis comes as a byproduct of such construction.

\subsection{A system of quantiles}
Consider a time series of $n$-dimensional random vectors, $Y_{t} \in \mathcal{Y} \subset \mathbb{R}^{n}$, and its natural filtration $\mathcal{F}_{t} = \sigma(\{ Y_{t},Y_{t-1},\ldots \})$, defined on the index set $\mathbb{Z}$. Then, there is a $k$-dimensional random vector $Z_t \in \mathcal{Z} \subset \mathbb{R}^{d}$ such that 
\begin{equation}
\mathbb{P}(Y_{t+h} \in A | \mathcal{F}_t) = \mathbb{P}(Y_{t+h} \in A | Z_t), \qquad \forall A \in \mathcal{F}_{t+h}
\end{equation}
for all $t\in\mathbb{Z}$ and $h\in\mathbb{Z}_{+}$.
Furthermore, we assume the time series is strictly stationary, which implies that the $h$ steps ahead forecasting distribution for the $i$th variable in the system can be defined as $Y_{t+h,i} | Z_{t}  = \bz_t \sim P_{h,i}( Y_{t+h,i} | \bz_{t})$, for any $i=1,\ldots,n$.
Therefore, conditioning on $Z_t = \bz_t$, the associated $\tau$th quantile, with $\tau \in (0,1)$, is given by:
\begin{equation}
Q_{i,h}^{\tau} \coloneqq Q_{i,h}^{\tau}(\bz_{t}) = \inf\left\{ \by \in \mathcal{Y}_{i}, \, P_{h,i}(\by|\bz_{t}) \leq \tau \right\},
\label{eq:Quantile_definition}
\end{equation}
where $\mathcal{Y}_{i}$ denotes the support of the $i$th variable in the system and captures the relationships between $\by$ and $\bz_{t}$.
In the following, we consider the quantile linear regression (QR) model as
\begin{equation}
Y_{t+h,i} = Q_{i,h}^{U}(\bz_{t}) = \bz_t' \bbeta_i^U,  \qquad  U\sim \mathcal{U}(0,1),
\label{eq:quantile}
\end{equation}
where $U$ and $\bz_t$ are independent and $\bbeta_i^U \in \R^d$ is a vector of coefficients.
Our approach focuses on the quantile regression of \cite{koenker1978regression} assuming exogeneity. When endogeneity is of concern, we suggest using instrumental variables (IV) to alleviate the issue based on the IVQR approach of \cite{chernozhukov2008instrumental}.

Based on the quantile regression model, Equation~\eqref{eq:Quantile_definition} implies that $Q_{i,h}^{\tau} = \bz_{t}' \bbeta_{i}^{\tau}$, where the coefficient vector can be estimated by solving the minimisation problem:
\begin{equation}
\widehat{\bbeta}_{ih}^{\tau} = \arg\min_{\bbeta} \sum_{t=1}^{T} \rho_\tau\big( y_{t+h,i} -\bz_{t}'\bbeta \big),
\end{equation}
where $\rho_\tau(x) = x(\tau - \I(x \leq 0))$ is the check loss function at quantile $\tau$ and $\I(\cdot)$ is the indicator function. The estimated values $\widehat{\bbeta}_{ih}^{\tau}$ are called regression quantiles and, given them, we can estimate the $\tau$th conditional quantile function of $y_{t+h,i}$ given $\bz_{t}$.
We remark that the use of the quadratic loss, $\rho_\tau(x) = \rho(x) = x^2$, results in the OLS estimator, whereas $\rho_\tau(x) = \rho(x) = |x|$ yields the least absolute deviation (LAD) estimator.

\subsection{Quantile Sensitivities}


Several studies have explored the dependence between different quantiles of various variables. One notable work in this area is the study by \cite{tobias2016covar}, which introduces the concept of Conditional Value at Risk (CoVaR). Given a system of time series, the CoVaR represents the Value at Risk (VaR) of one variable conditional on the event that another variable reaches its VaR. This approach provides insights into the tail dependence and extreme risk associated with the joint behaviour of different variables.
In particular, this concept is interesting because it elicits the riskiness of one variable in the future, given that another variable is already at risk.
Precisely translated into our notation, the CoVaR of the $h$ steps ahead $\tau$th quantile of the $i$th variable given the one step ahead $\tau'$th quantile of the $j$th variable defined as:
\begin{equation}
\mathbb{P}\big( Y_{t+h,i} \leq \text{CoVaR}_{h,1}^{\tau,\tau'} | Z_{t}, \, Y_{t+1,j} = Q_{j,1}^{\tau'} \big) = \tau.
\end{equation}
Another attempt is made by \cite{lee2021impulse}, who formulate the problem in a VAR structure. Building on the VAR literature, they propose a quantile impulse response function (QIRF) as the change of the $\tau$th quantile of the $i$th variable in response to an exogenous shock to the system, that is:
\begin{equation}
\text{QIRF}_i^\tau(h) \coloneqq \frac{\partial Q_{i,h}^{\tau}}{\partial \bepsilon_{t}},
\end{equation}
where $\bepsilon_{t}$ is the vector of mean zero disturbances with a diagonal covariance matrix decomposed from the standard mean regression structural VAR. Instead, \cite{Manganelli2021quantileIRF} propose a different quantile impulse response function to perform stress tests in a structural quantile VAR (QVAR) model that captures nonlinear relationships among macroeconomic variables. Finally, \cite{Han2022vq} propose several ways of constructing and estimating quantile impulse response functions through local projection methods.

Our approach differs from theirs by accounting for the potential variation in the impact of different quantiles of money growth on inflation.
In particular, we combine the approaches of \cite{tobias2016covar} and \cite{lee2021impulse} and define the quantile sensitivity as the responsiveness of the $h$ steps ahead quantile of a variable of interest to the change of the 1 step ahead quantile of another variable.

\begin{definition}[Quantile Sensitivity] \label{def:QS_definition}
For an $n$-dimensional time series process $Y_t$, $n > 1$, the quantile sensitivity (\QS) of the $h$-steps ahead $\tau$ quantile of the $i$th variable with respect to a change of the $1$-step ahead $\tau'$ quantile of the $j$th variable is defined as:
\begin{equation}
\QS_{i,j,h,1}^{\tau,\tau'} \equiv \QS_{i,j}(h;\tau,\tau') \coloneqq \frac{\partial Q_{i,h}^{\tau}}{\partial Q_{j,1}^{\tau'}}.
\label{eq:QS_definition}
\end{equation}
\end{definition}

Consider the relationship between $\widetilde{Q}_{i,h}^\tau$ and $\widetilde{Q}_{j,1}^{\tau'}$ as represented by $\widetilde{Q}_{i,h}^\tau = f\big( \widetilde{Q}_{j,1}^{\tau'} \big)$.
Then, using a first-order Taylor approximation of $f(\cdot)$ about $Q_{j,1}^{\tau'}$ yields:
\begin{align}
\notag
f\big( \widetilde{Q}_{j,1}^{\tau'} \big) & \approx f\big( Q_{j,1}^{\tau'} \big) + \frac{\partial f}{\partial \widetilde{Q}_{j,1}^{\tau'}} \Bigg|_{\widetilde{Q}_{j,1}^{\tau'} = Q_{j,1}^{\tau'}} \times \big( \widetilde{Q}_{j,1}^{\tau'} - Q_{j,1}^{\tau'} \big), \\
\notag 
\widetilde{Q}_{i,h}^\tau & \approx Q_{i,h}^{\tau} + \frac{\partial Q_{i,h}^\tau}{\partial Q_{j,1}^{\tau'}} \times \left( \widetilde{Q}_{j,1}^{\tau'} - Q_{j,1}^{\tau'} \right) \\
&\approx Q_{i,h}^{\tau} + { \QS_{i,j}(h;\tau,\tau') } \times \left( \widetilde{Q}_{j,1}^{\tau'} - Q_{j,1}^{\tau'} \right),
\label{eq:QS_Taylor}
\end{align}
which implies that the change of the $\tau$th quantile of variable $i$, $\widetilde{Q}_{i,h}^{\tau} - Q_{i,h}^{\tau}$, in response to a change the $\tau'$th quantile of variable $j$, $\widetilde{Q}_{j,1}^{\tau'} - Q_{j,1}^{\tau'}$, are proportional, with a proportionality factor given by the quantile sensitivity \QS.

Moreover, by considering all $\tau\in (0,1)$, we can assess the relationship between perturbations in the $\tau'$th quantile of variable $j$ and the resulting changes in the entire $h$ steps ahead distribution of variable $i$. This allows for a comprehensive analysis of how different quantiles of one variable are associated with variations in the distribution of another variable over multiple steps ahead.

\subsection{Estimation of the Quantile Sensitivities}  \label{sec:estimation_QS}
The quantile sensitivity defined in Equation~\eqref{eq:QS_definition} involves expressing $Q_{i,h}^\tau$ as a function of $Q_{j,1}^{\tau'}$. In this section, we demonstrate the construction of this relationship using linear quantile regressions. By utilizing linear quantile regression, it is possible to estimate the conditional quantiles of variable $i$ given different quantiles of variable $j$, thus allowing to quantify the sensitivity of $Q_{i,h}^\tau$ to changes in $Q_{j,1}^{\tau'}$.

Without the loss of generality, we consider a set of quantile levels $\boldsymbol{\tau} = \{ \tau_1,\ldots,\tau_k \}$, $k > 1$. Let $\bq_{j,1} = \big( Q_{j,1}^{\tau_1},\ldots,Q_{j,1}^{\tau_k} \big)' \in \mathbb{R}^{k}$ be a $k$-dimensional vector of quantiles for variable $j$, with $j \in \{ 1,\ldots,n\}$. Assuming a homogeneous model across all quantile levels, we have $\bq_{j,1} = \bB_{j,1} \bz_t$, where $\bB_{j,1} = (\bbeta_{j1}^{\tau_1},\ldots,\bbeta_{j1}^{\tau_k})' \in \mathbb{R}^{k\times d}$ denote the matrix-variate quantile coefficients.
We can now construct the system of quantiles of the one-step ahead predictive distribution as:
\begin{equation}
\bQ_{1} = ( \bq_{11}', \bq_{21}', \ldots, \bq_{n1}' )' = \bB_1 \bz_t,
\end{equation}
with $\bB_1 = (\bB_{11}', \bB_{21}', \ldots, \bB_{n1}')' \in \mathbb{R}^{nk\times d}$.
Assuming that $\bB_1' \bB_1$ is invertible\footnote{Note that $\beta_{i}^U=\frac{\partial}{\partial z}Y_{t+h,i}$. Thus, this matrix asymptotically converges to $\mathbb{E}\left[ \frac{\partial}{\partial z}Y_{t+h,i} \left(\frac{\partial}{\partial z}Y_{t+h,i}\right)'\right].$}, we have
\begin{equation}
\bz_t = \left( \bB_1' \bB_1 \right)^{-1} \bB_1' \bQ_1.
\end{equation}
A similar construction can be obtained for the collection of $h$-steps ahead conditional quantiles $\bQ_h = \bB_h \bz_t = \bB_h \left( \bB_1' \bB_1 \right)^{-1} \bB_1' \bQ_1$.
Thus, Equation~\eqref{eq:QS_definition} can be obtained by selecting elements from the $(nk \times nk)$ matrix:
\begin{equation}
\overline{\bB}_{h,1} = \bB_h \left( \bB_1' \bB_1 \right)^{-1} \bB_1'. 
\end{equation}
In this formulation, it is important to note that the linear projection approach assumes time-homogeneous dependence. This implies that the matrix describing the interlink between quantiles does not depend on $\bz_t$ and is supposed to be constant over time.
Therefore, the linear construction utilised in this article assumes a static relationship between the variables and does not consider potential time-varying dynamics in the dependence structure between the variables.

The static dependence implied by the linearity assumption still allows us to investigate how the distribution of inflation changes in response to a change in money growth at different periods.
This is because, given the set of quantile regression coefficients, we can find the quantile level of the observed money growth at time $t$, that is $\tau_t = \inf \{ \tau \in (0,1) : \bz_t' \bbeta_{i,\tau} < y_{t,j} \}$.
Then
\begin{equation}
    \QS_{i,j}(h;\tau,\tau_t) = \frac{\partial Q_{i,h}^\tau}{\partial Q_{j,1}^{\tau_t}}
\end{equation}
gives the responsiveness of the quantiles at $t+h$ of any other variable $i$ to a change at time $t+1$ of the money supply from its current level.

\section{Empirical Application}  \label{sec:results}

In this real-data application, we investigate the response of change in money growth to inflation. Therefore, we set $Y_{i,t+h}$ as the inflation measure and $\bZ_t$ as the matrix of covariates, comprising an intercept term and lagged values of both money growth and inflation.
As for the forecast horizons, we focus on the 12 months (or 1-year) ahead forecast, thus setting $h=12$.
Moreover, to deeper the analysis of the relationship between these key macroeconomic variables, we apply the proposed \QS\ measure to both US and European data to identify common behaviours and idiosyncratic features.
The US PCE and core inflation data were collected from the St. Louis Fred database, while the Euro Area HICP and core inflation data were obtained from the ECB statistical data warehouse. We refer to the Supplement for additional details.


The rest of the section reports and discusses the main findings for the two economies.
First, we examine the short-run (1-year) impact of money growth on the distribution of inflation and its disaggregated measures.
Second, we assess the existence of potential temporal variation in the distributional effects of money growth on inflation.
%
All the empirical analyses have been performed by assessing the impact of both extreme shocks to money growth ($\tau^M = 0.10$ and $\tau^M = 0.90$) and ``ordinary'' or changes to the median ($\tau^M = 0.50$). For each of them, we investigated its impact on the entire distribution of the inflation variable, as approximated by the fine grid of quantiles $\tau^I \in \{ 0.01, 0.02, \ldots, 0.99 \}$.

\begin{figure}[t!h]
\centering
\tiny
\begin{tabular}{c  c c c c c c}
 & PCE & Core PCE & PCE Durables & PCE Nondurables & PCE Services & PCE Energy \\
\rotatebox{90}{\hspace*{16pt} \centering 10\%} \hspace*{-10pt} &
\includegraphics[scale=0.19]{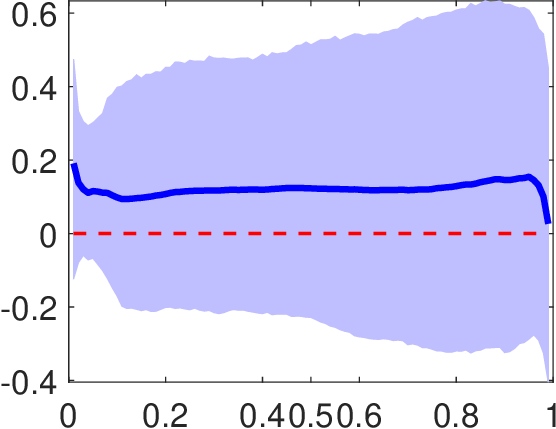} &
\includegraphics[scale=0.19]{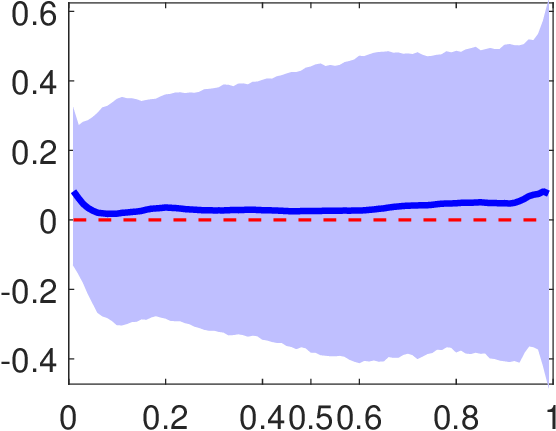} &
\includegraphics[scale=0.19]{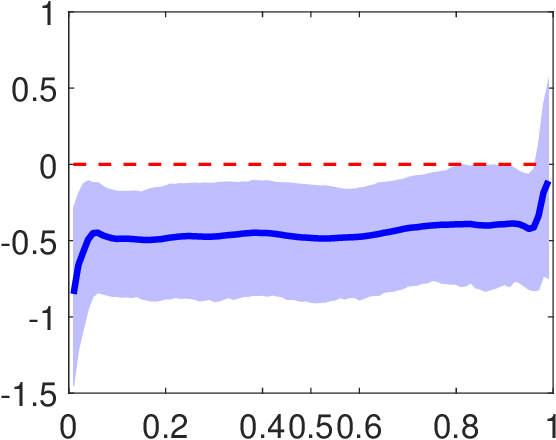} &
\includegraphics[scale=0.19]{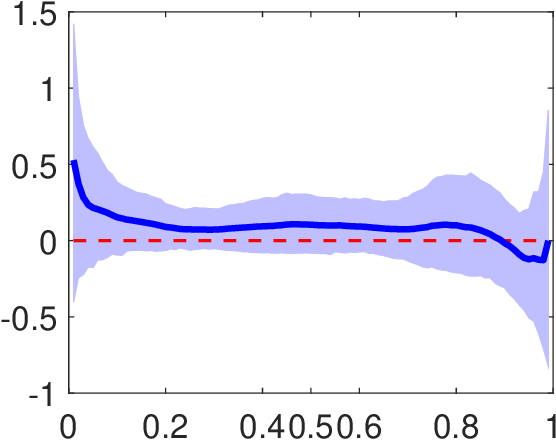} &
\includegraphics[scale=0.19]{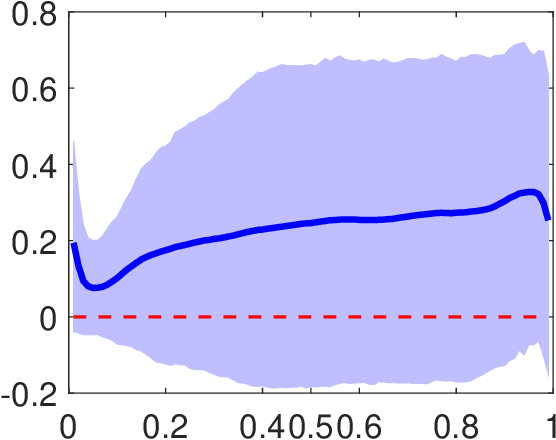} &
\includegraphics[scale=0.19] {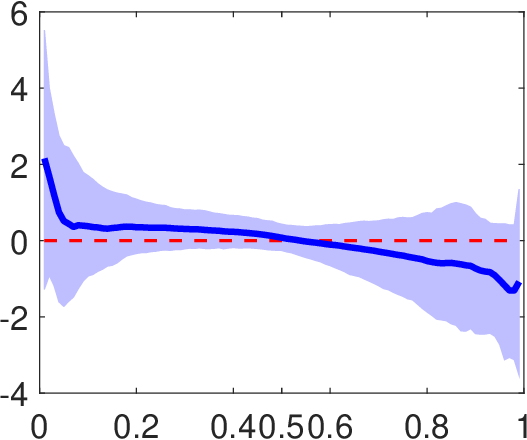} \\[15pt]
\rotatebox{90}{\hspace*{16pt} \centering 50\%} \hspace*{-10pt} &
\includegraphics[scale=0.19]{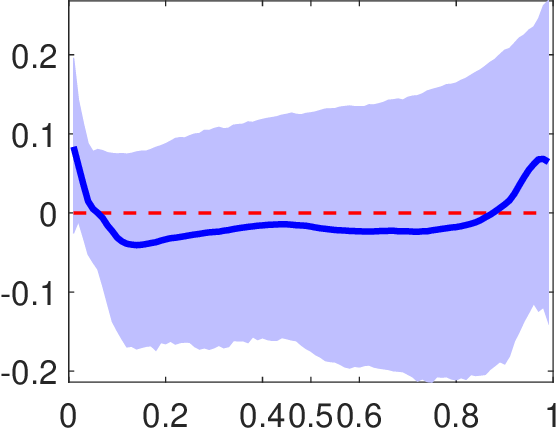} &
\includegraphics[scale=0.19]{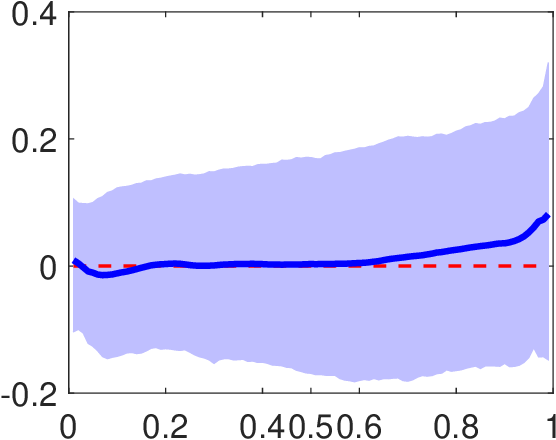} &
\includegraphics[scale=0.19]{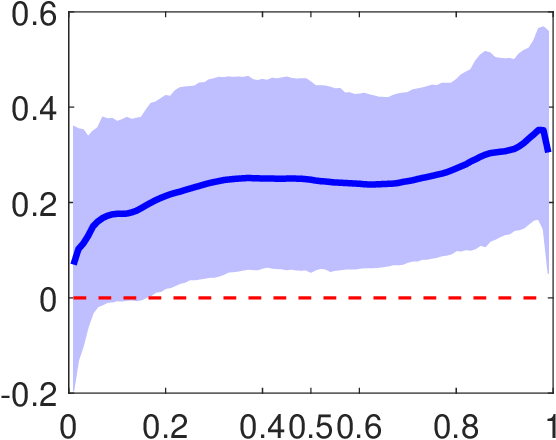} &
\includegraphics[scale=0.19]{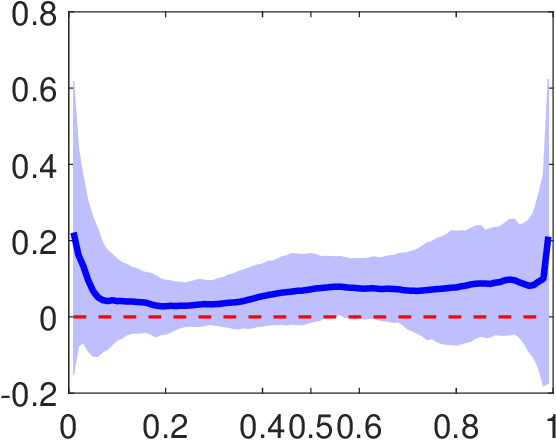} &
\includegraphics[scale=0.19]{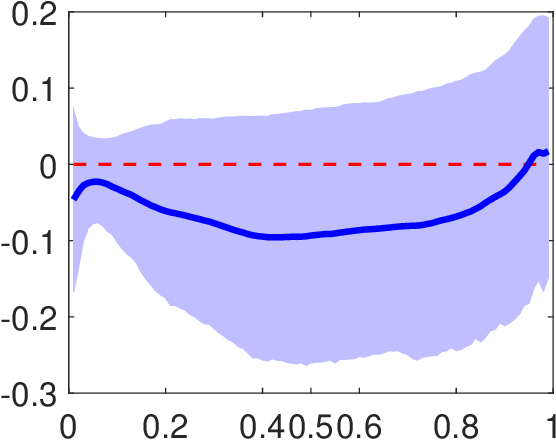} &
\includegraphics[scale=0.19]{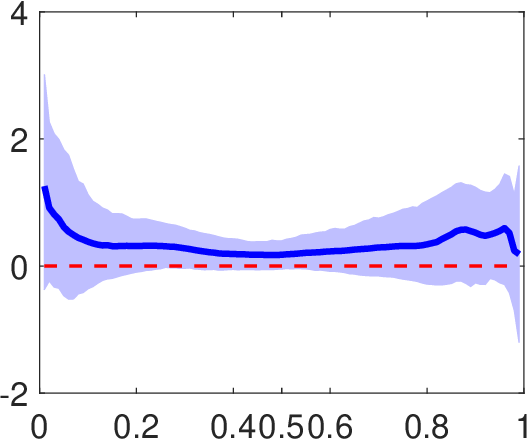} \\[15pt]
\rotatebox{90}{\hspace*{16pt} \centering 90\%} \hspace*{-10pt} &
\includegraphics[scale=0.19]{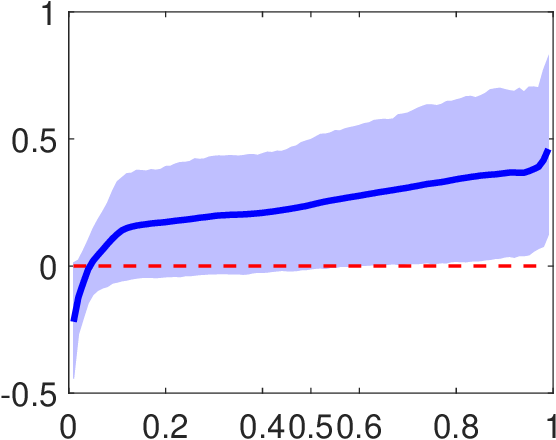} &
\includegraphics[scale=0.19]{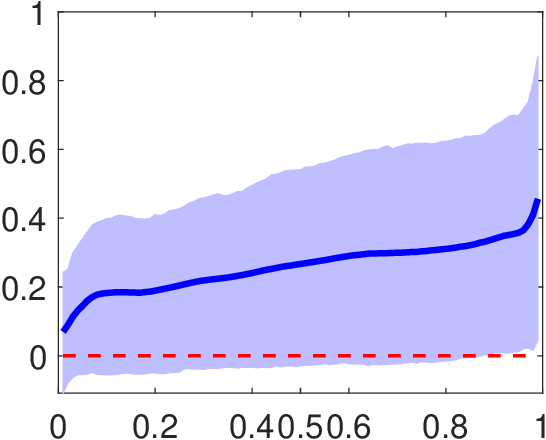} &
\includegraphics[scale=0.19]{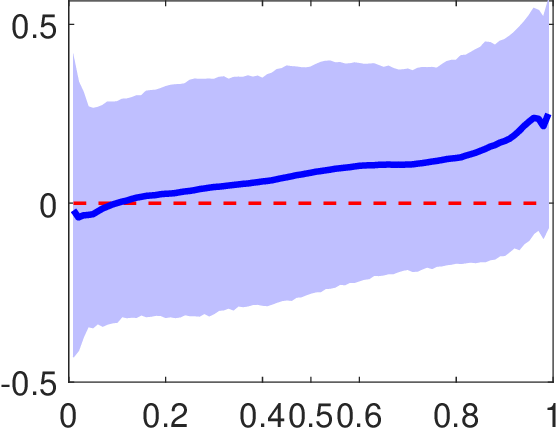} &
\includegraphics[scale=0.19]{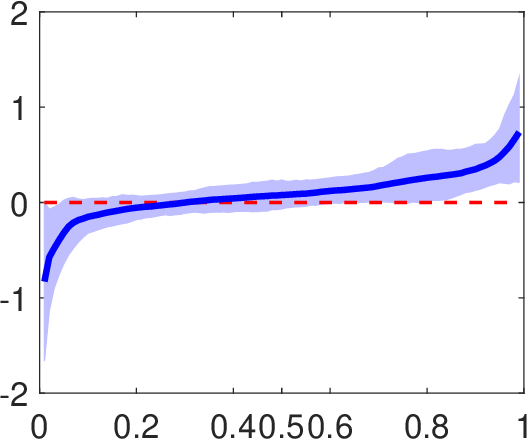} &
\includegraphics[scale=0.19]{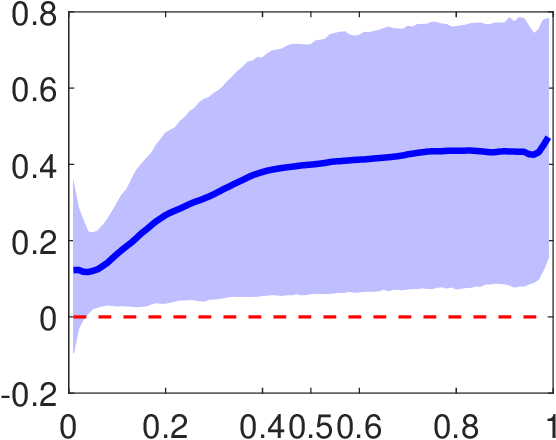} &
\includegraphics[scale=0.19]{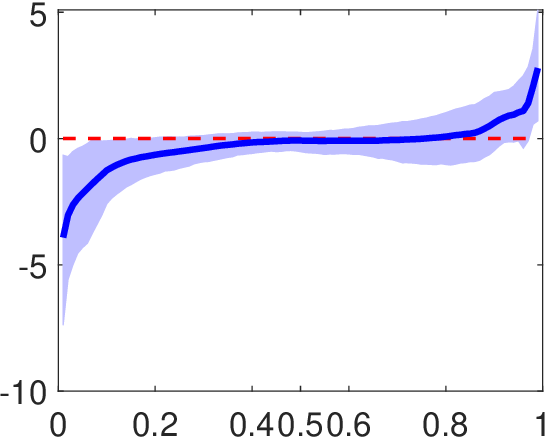}
\end{tabular}
\caption{Distributional impact of the 1-year ahead impact of money growth on US PCE inflation across selected quantiles $\tau^M \in \{ 0.10,0.50,0.90 \}$.
Notes: the thick blue line represents the mean quantile responses of the 1-year ahead impact of money growth on US PCE inflation. The shaded blue area represents the 68\% bootstrapped confidence interval.}
\label{fig:USQIRF}
\end{figure}

Figure~\ref{fig:USQIRF} presents the distributional effect of the 1-year ahead money growth impact on US PCE inflation, core inflation, and its four major disaggregated measures across selected quantiles. The findings reveal that at the 10th quantile, money growth significantly negatively impacts the entire distribution of durable goods inflation (i.e., for every level of $\tau^I$).
However, a change in the lower quantile of money growth does not have a significant distributional impact on the other three disaggregated measures and the overall PCE inflation. This is mainly driven by the high uncertainty of the (positive but not significant) effect on services inflation.
Turning to the upper tail of money growth, specifically the 90th percentile, a significant positive impact is observed on the right tail of the overall PCE inflation. This effect leads to an increase in the thickness of the distribution tail and a rightward skewness. Furthermore, this rightward skewness contributes to the asymmetry of the PCE inflation distribution.
Regarding the disaggregated measures, shocks to the upper tail of money growth also result in thicker-tailed but symmetric distributions for both PCE non-durable goods and energy (goods and services) inflation. Instead, we find evidence of a significant rightward shift in the distribution of PCE service inflation. However, the upper quantile of money growth does not significantly affect PCE durable goods inflation.
Summarizing, these findings suggest that the increasing positive skewness observed in the overall PCE inflation distribution due to changes in the upper quantile of money growth is mainly driven by the shift in PCE service inflation towards the right.

\begin{figure}[t!h]
\centering
\tiny
\hspace*{-0.8cm}
\begin{tabular}{c  c c c c c c c}
 & HICP & Core HICP & HICP Unproc food & HICP Proc food & HICP Ind goods food & HICP Services & HICP Energy \\
\rotatebox{90}{\hspace*{16pt} \centering 10\%} \hspace*{-10pt} &
\includegraphics[scale=0.19]{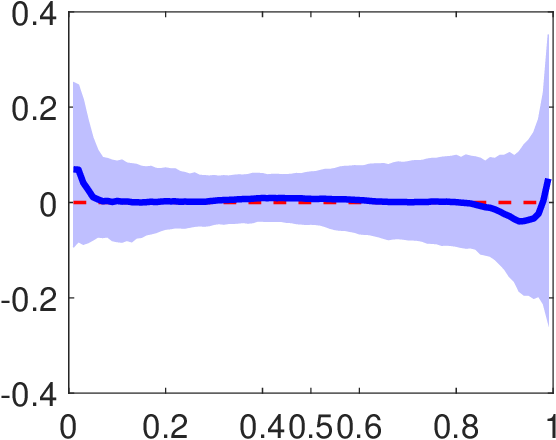} &
\includegraphics[scale=0.19]{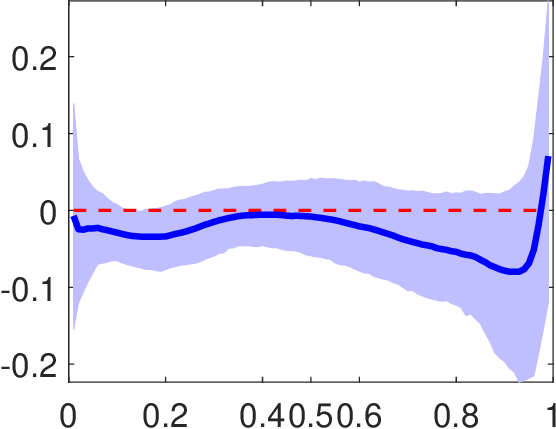} &
\includegraphics[scale=0.19]{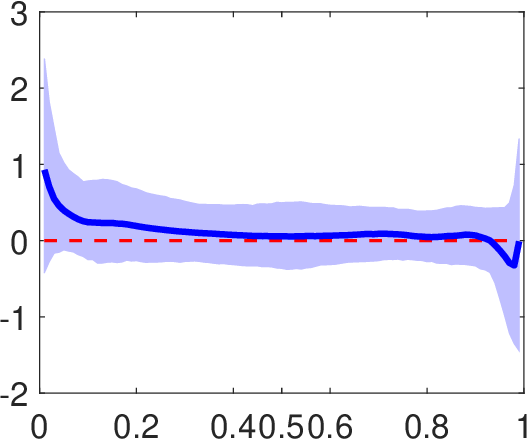} &
\includegraphics[scale=0.19]{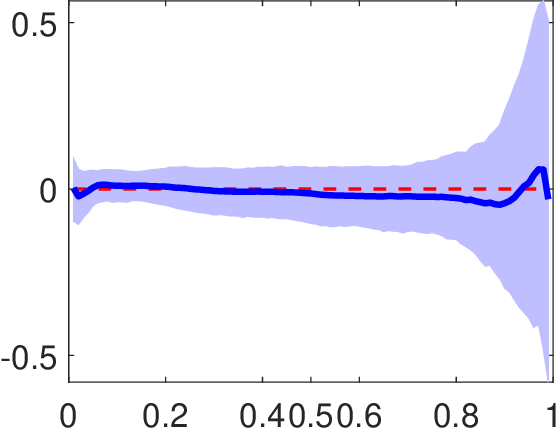} &
\includegraphics[scale=0.19]{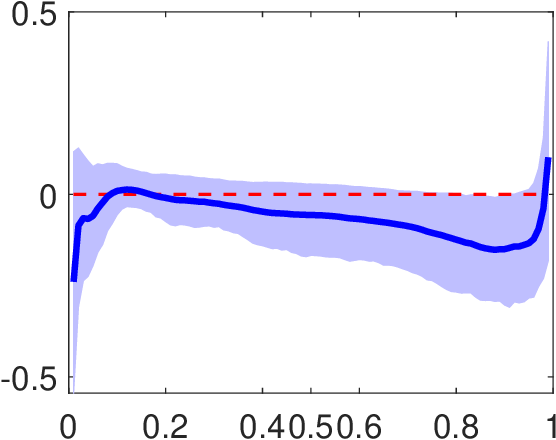} &
\includegraphics[scale=0.19]{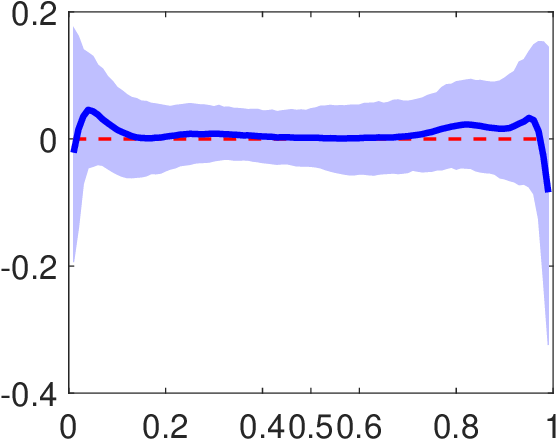} &
\includegraphics[scale=0.19]{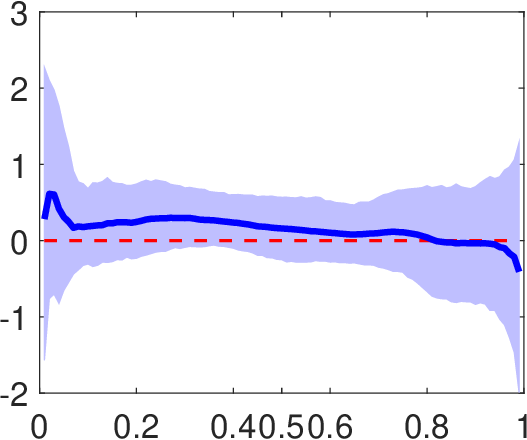} \\[15pt]
\rotatebox{90}{\hspace*{16pt} \centering 50\%} \hspace*{-10pt} &
\includegraphics[scale=0.19]{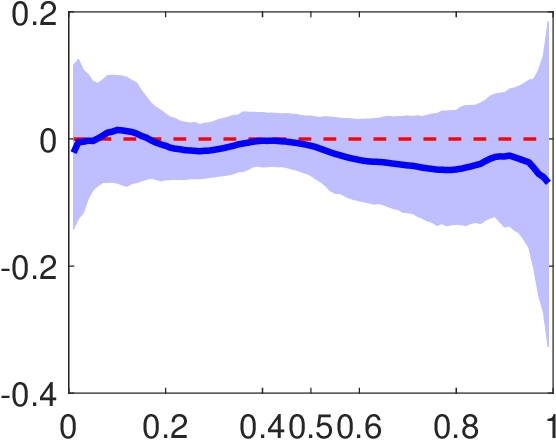} &
\includegraphics[scale=0.19]{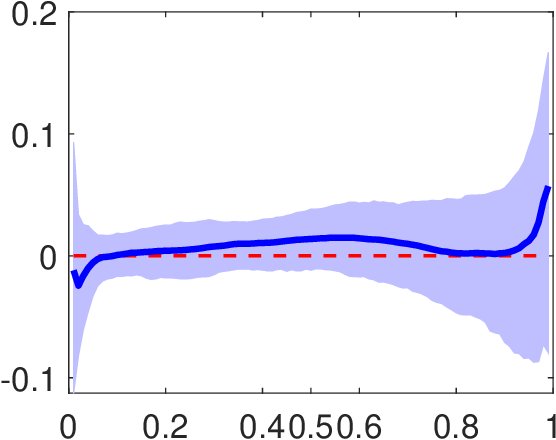} &
\includegraphics[scale=0.19]{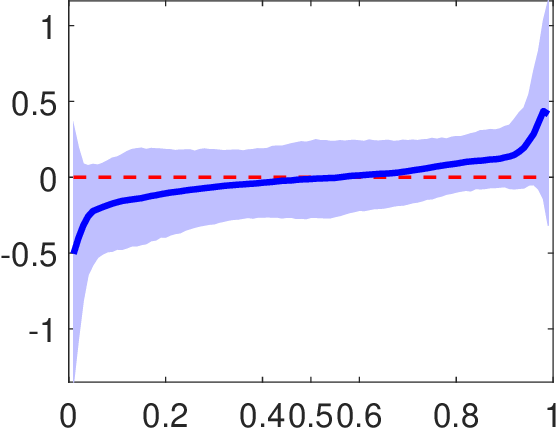} &
\includegraphics[scale=0.19]{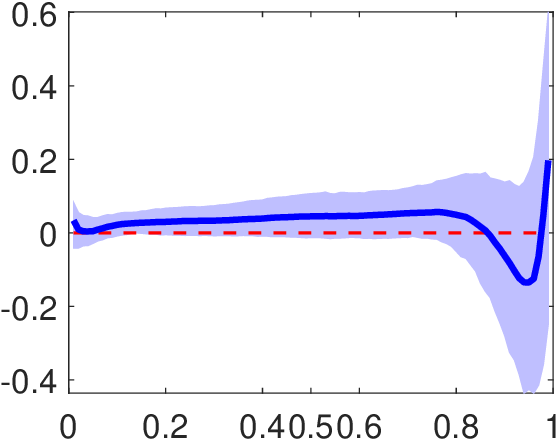} &
\includegraphics[scale=0.19]{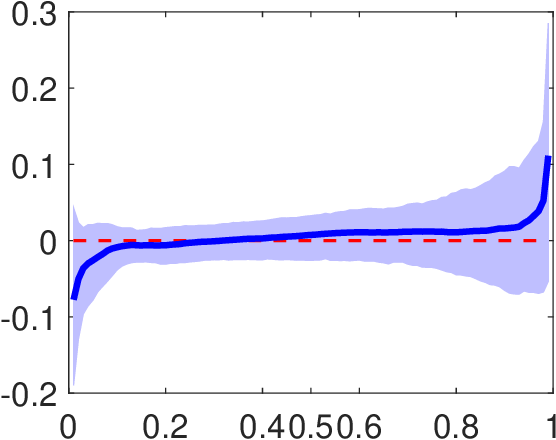} &
\includegraphics[scale=0.19]{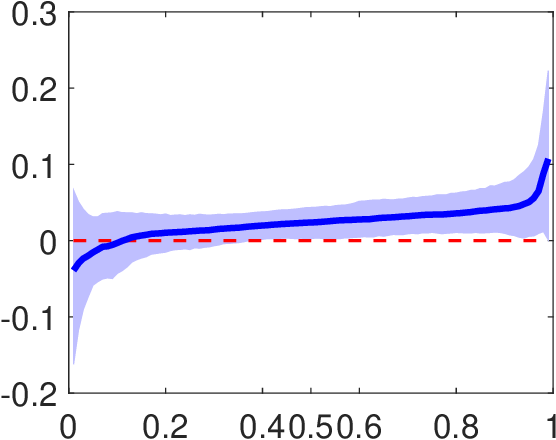} &
\includegraphics[scale=0.19]{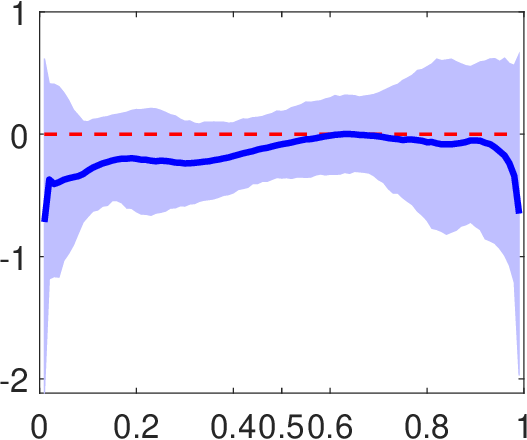} \\[15pt]
\rotatebox{90}{\hspace*{16pt} \centering 90\%} \hspace*{-10pt} &
\includegraphics[scale=0.19]{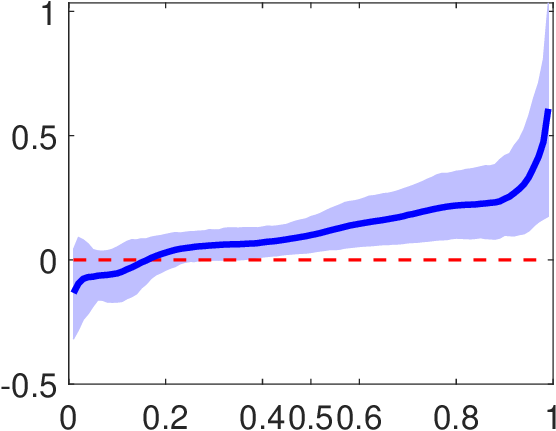} &
\includegraphics[scale=0.19]{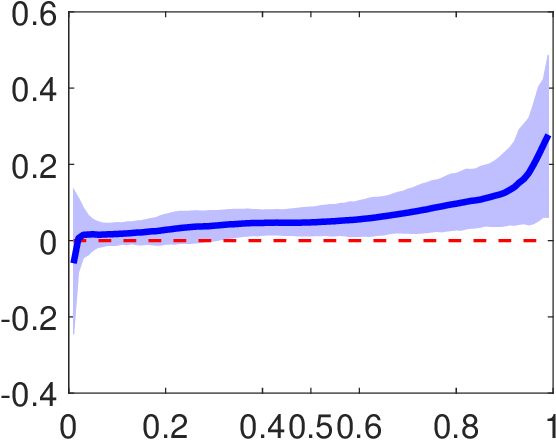} &
\includegraphics[scale=0.19]{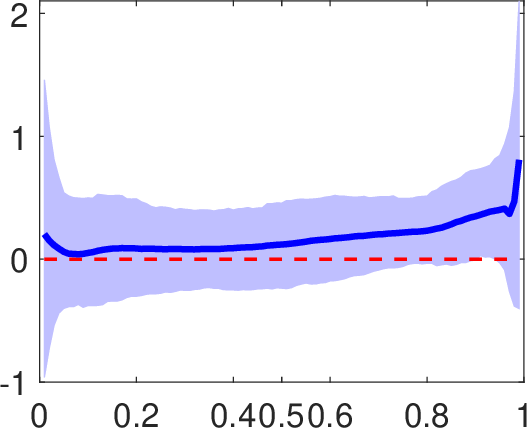} &
\includegraphics[scale=0.19]{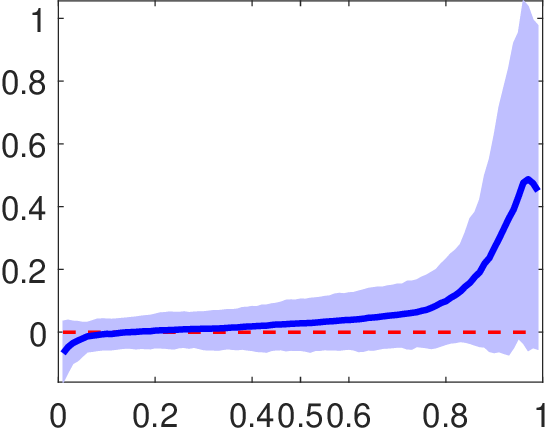} &
\includegraphics[scale=0.19]{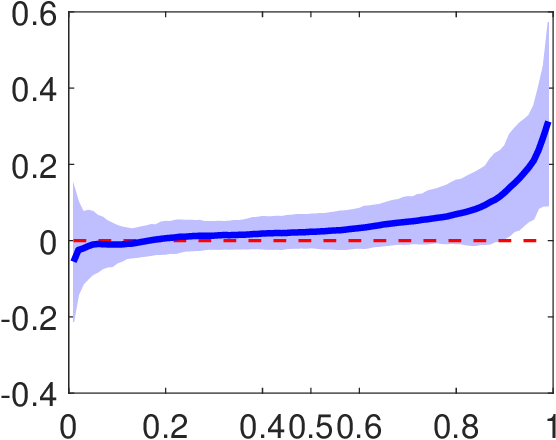} &
\includegraphics[scale=0.19]{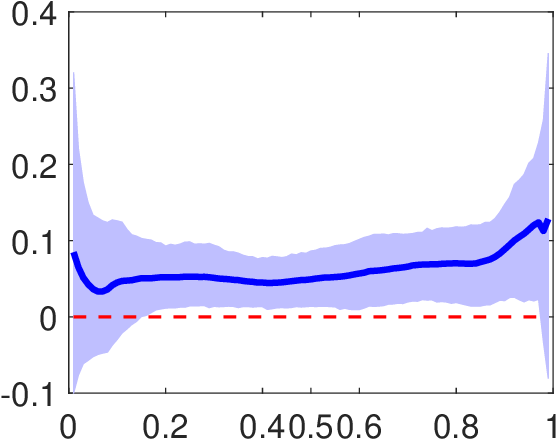} &
\includegraphics[scale=0.19]{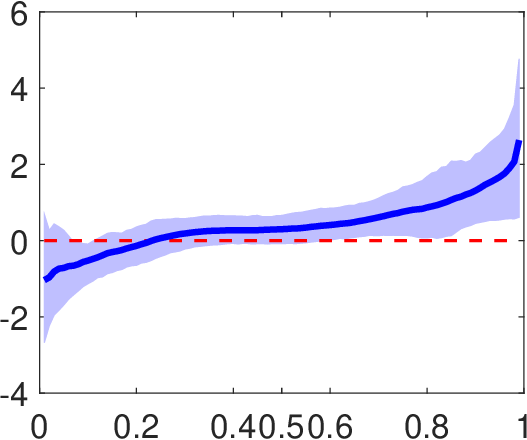}
\end{tabular}
\caption{Distributional impact of the 1-year ahead impact of money growth on Euro HICP inflation across selected quantiles $\tau^M = \{ 0.10,0.50,0.90 \}$.
Notes: the thick blue line represents the mean quantile responses of the 1-year ahead impact of money growth on Euro HICP inflation. The shaded blue area represents the 68\% bootstrapped confidence interval.}
\label{fig:EUQIRF}
\end{figure}

We also present the distributional effect of the 1-year ahead money growth impact on Euro area HICP inflation, core inflation, and its five major disaggregated measures across selected quantiles in Figure~\ref{fig:EUQIRF}.
Similarly to the US case, our findings indicate that at the lower quantile (specifically, the 10th percentile), money growth does not significantly impact HICP inflation and its associated disaggregated measures.
However, for the upper quantile of money growth, we find a discernible shift of the rightmost part of the distribution of overall HICP inflation, manifesting as right skewness. This emerges from the positive impact on all the quantiles $\tau^I > 0.45$. Moreover, these extreme changes to the right tail of money growth are found to increase in magnitude along the quantiles of the HICP distribution, thus resulting in a significant right-skewing effect. Overall, this shift contributes to an increase in the thickness of the tails and an asymmetry in the distribution.
Notably, the aforementioned distributional shift appears to be primarily driven by HICP services and energy inflation: the former seems responsible for the level shift (to the right) of the entire HICP distribution, whereas the latter accounts for the increased right-skewness.
On the other hand, the upper quantile of money growth does not exert a significant distributional impact on HICP inflation is related to unprocessed, processed, and industrial goods.

Consequently, the empirical results for both the US and the Euro area suggest that, at the upper quantile of money growth (or in the presence of excess money growth), the distribution of inflation becomes more positively skewed, with services playing a crucial role in this distributional shift, followed by the energy sector and durable goods.
Conversely, at the 10th percentile and the median of money growth, we find no substantial distributional impact on inflation and its disaggregated measures in the US and the Euro area.\footnote{We also report the distributional impact of the 1-year ahead impact of money growth for the 25\% and 75\% quantiles in the supplementary material. These results display very similar dynamics to the 10\% and 90\% quantiles reported in the US and Euro Area.}
%

Our empirical findings appear to align with the conclusions presented in \cite{Schnabel2023}, suggesting that the impact of monetary expansion on inflation is closely tied to the prevailing economic conditions. To illustrate, in the aftermath of the Great Recession, both central banks in their respective economies implemented a quantitative easing (QE) policy, which involved substantial asset acquisitions. However, it is noteworthy that during this period, financial markets experienced high illiquidity. As highlighted in \cite{Schnabel2023}, these extensive asset purchases had a minimal effect on money growth, and consequently, they did not significantly impact inflation. This particular scenario sheds light on the relatively minor effect on inflation observed within the lower to median quantiles of money growth.


In contrast, during the post-pandemic era, \cite{Schnabel2023} contends that households showed diminished sensitivity to price increases, and the excess money growth resulting from fiscal measures could perpetuate detrimental cost-push shocks. Specifically, robust money growth can exacerbate the persistence of such supply-side shocks. This perspective aligns with the findings of \cite{golosov2007menu}, who argue that menu costs predominantly result in individual prices remaining stable for extended periods, undergoing significant adjustments only in response to substantial idiosyncratic shocks within the economy. Our empirical findings seemingly support both \cite{Schnabel2023} and \cite{golosov2007menu} assertions, revealing that within the upper quantile of money growth, the distribution of inflation skews towards a more positive trajectory.

Furthermore, our empirical findings also align with \cite{blanco2022nonlinear} research, which provides compelling evidence that underscores the substantial influence of the frequency of price changes on the dynamics of prices, particularly during periods of high inflation. Their study demonstrates that a significant monetary shock exerts a more than doubling effect on the frequency of price adjustments within the economy. This suggests that monetary (money) shocks have a pronounced amplifying impact on the frequency of price changes when inflation levels are elevated.


Finally, all the previous empirical studies examining the relationship between money growth and inflation have been examined through the lens of the conditional mean of the distribution \citep[see][]{grauwe2005inflation,gertler2018monetary,borio2023does} and each of the studies has yielded contradictory results. To comprehensively understand how money growth influences inflation, it is essential to consider the broader distributional effects within a standard macroeconomic model. However, this exploration is reserved for future research avenues.

To better understand the distributional impact of the 1-year ahead effect of money growth on US PCE inflation and Euro area HICP inflation, we report in the Supplement a 3D plot illustrating this relationship across all quantiles.
Besides, we apply the quantile sensitivity method over specific periods and find evidence of temporal variation in the distributional impact of money growth on inflation and its disaggregate measures between the US and the Euro area (see Supplement).

\section{Conclusions}  \label{sec:conclusion}
We proposed an innovative framework for examining the short-run impact of money growth on inflation and its disaggregate measures in the US and the Euro area. Our framework focuses on quantile dependence, providing a novel approach to analyze the distributional effects. This aspect of the relationship between money growth and inflation has not been previously explored in the existing literature.

Through empirical analysis of both economies, we found evidence that variations in the upper quantile of the money growth distribution significantly influence the distribution of inflation and its disaggregate measures, affecting both its level and skewness. Conversely, the lower and median quantiles of the money growth distribution exhibit a relatively negligible effect on the distribution of inflation. Our findings are consistent with \cite{golosov2007menu} and \cite{blanco2022nonlinear}, where menu cost plays a dominant role in keeping individual prices constant for most of the time and large money shocks have more than a doubling effect on the frequency of price changes.

Our research emphasizes the importance of incorporating money growth distributional effects into macroeconomic models. This detailed understanding is crucial for macroeconomists aiming to comprehend the fundamental mechanisms linking money growth and inflation.
The proposed method is versatile and can be applied to investigate various other macroeconomic or financial phenomena, such as to explore the relationship between oil prices and stock returns or a growth-at-risk scenario.

\bibliographystyle{chicago}
\bibliography{biblio}

\begin{thebibliography}{}

\bibitem[\protect\citeauthoryear{Benati}{Benati}{2009}]{benati2009long}
Benati, L. (2009).
\newblock Long run evidence on money growth and inflation.
\newblock Technical Report 1027, ECB working paper.

\bibitem[\protect\citeauthoryear{Blanco, Boar, Jones, and Midrigan}{Blanco
  et~al.}{2022}]{blanco2022nonlinear}
Blanco, A., C.~Boar, C.~Jones, and V.~Midrigan (2022).
\newblock Nonlinear inflation dynamics in menu cost economies.
\newblock Technical report.

\bibitem[\protect\citeauthoryear{Borio, Hofmann, and Zakraj{\v{s}}ek}{Borio
  et~al.}{2023}]{borio2023does}
Borio, C., B.~Hofmann, and E.~Zakraj{\v{s}}ek (2023).
\newblock Does money growth help explain the recent inflation surge?
\newblock Technical report, Bank for International Settlements.

\bibitem[\protect\citeauthoryear{Chavleishvili and Manganelli}{Chavleishvili
  and Manganelli}{2021}]{Manganelli2021quantileIRF}
Chavleishvili, S. and S.~Manganelli (2021).
\newblock Forecasting and stress testing with quantile vector autoregression.
\newblock Technical Report 2330, ECB Working Paper.

\bibitem[\protect\citeauthoryear{Chernozhukov and Hansen}{Chernozhukov and
  Hansen}{2008}]{chernozhukov2008instrumental}
Chernozhukov, V. and C.~Hansen (2008).
\newblock Instrumental variable quantile regression: A robust inference
  approach.
\newblock {\em Journal of Econometrics\/}~{\em 142\/}(1), 379--398.

\bibitem[\protect\citeauthoryear{Gertler and Hofmann}{Gertler and
  Hofmann}{2018}]{gertler2018monetary}
Gertler, P. and B.~Hofmann (2018).
\newblock Monetary facts revisited.
\newblock {\em Journal of International Money and Finance\/}~{\em 86},
  154--170.

\bibitem[\protect\citeauthoryear{Golosov and Lucas}{Golosov and
  Lucas}{2007}]{golosov2007menu}
Golosov, M. and R.~E. Lucas (2007).
\newblock Menu costs and {Phillips} curves.
\newblock {\em Journal of Political Economy\/}~{\em 115\/}(2), 171--199.

\bibitem[\protect\citeauthoryear{Grauwe and Polan}{Grauwe and
  Polan}{2005}]{grauwe2005inflation}
Grauwe, P.~D. and M.~Polan (2005).
\newblock Is inflation always and everywhere a monetary phenomenon?
\newblock {\em Scandinavian Journal of Economics\/}~{\em 107\/}(2), 239--259.

\bibitem[\protect\citeauthoryear{Han, Jung, and Lee}{Han
  et~al.}{2022}]{Han2022vq}
Han, H., W.~Jung, and J.~H. Lee (2022).
\newblock Estimation and inference of quantile impulse response functions by
  local projections: With applications to var dynamics.
\newblock {\em Journal of Financial Econometrics\/}, nbac026.

\bibitem[\protect\citeauthoryear{King}{King}{2022}]{king2022monetary}
King, M. (2022).
\newblock Monetary policy in a world of radical uncertainty.
\newblock {\em Economic Affairs\/}~{\em 42\/}(1), 2--12.

\bibitem[\protect\citeauthoryear{Koenker and Bassett}{Koenker and
  Bassett}{1978}]{koenker1978regression}
Koenker, R. and G.~Bassett (1978).
\newblock Regression quantiles.
\newblock {\em Econometrica\/}, 33--50.

\bibitem[\protect\citeauthoryear{Laidler}{Laidler}{2021}]{laidler2021personal}
Laidler, D. (2021).
\newblock A personal view from the wrong side of the subsequent fifty years.
\newblock Technical report, University of Western Ontario, Department of
  Economics.

\bibitem[\protect\citeauthoryear{Lee, Kim, and Mizen}{Lee
  et~al.}{2021}]{lee2021impulse}
Lee, D.~J., T.-H. Kim, and P.~Mizen (2021).
\newblock Impulse response analysis in conditional quantile models with an
  application to monetary policy.
\newblock {\em Journal of Economic Dynamics and Control\/}~{\em 127}, 104102.

\bibitem[\protect\citeauthoryear{Lucas}{Lucas}{1980}]{lucas1980two}
Lucas, R.~E. (1980).
\newblock Two illustrations of the quantity theory of money.
\newblock {\em The American Economic Review\/}~{\em 70\/}(5), 1005--1014.

\bibitem[\protect\citeauthoryear{Sargent and Surico}{Sargent and
  Surico}{2011}]{sargent2008monetary}
Sargent, T.~J. and P.~Surico (2011).
\newblock Two illustrations of the quantity theory of money: Breakdowns and
  revivals.
\newblock {\em American Economic Review\/}~{\em 101\/}(1), 109 -- 128.

\bibitem[\protect\citeauthoryear{Schnabel}{Schnabel}{2023}]{Schnabel2023}
Schnabel, I. (2023).
\newblock Money and inflation.
\newblock Thünen Lecture by Isabel Schnabel, Member of the Executive Board of
  the ECB, at the annual conference of the Verein für Socialpolitik.

\bibitem[\protect\citeauthoryear{Tobias and Brunnermeier}{Tobias and
  Brunnermeier}{2016}]{tobias2016covar}
Tobias, A. and M.~K. Brunnermeier (2016).
\newblock {CoVaR}.
\newblock {\em American Economic Review\/}~{\em 106\/}(7), 1705.

\end{thebibliography}


\end{document}